\documentclass[final,5p,times,twocolumn]{elsarticle}

\usepackage{amssymb}

\usepackage{lineno}

\journal{Nuclear Instruments and Methods in Physics Research Section A}

\begin{document}

\begin{frontmatter}



\title{The quality assurance test of the SliT ASIC for
the J-PARC muon $g-2$/EDM experiment}


\author[artsci_kyushu]{Takashi Yamanaka}
\author[kek]{Yoichi Fujita}
\author[kek]{Eitaro Hamada}
\author[kek]{Tetsuichi Kishishita}
\author[kek]{Tsutomu Mibe}
\author[niigata]{Yutaro Sato}
\author[toyama]{Yoshiaki Seino}
\author[kek]{Masayoshi Shoji}
\author[phys_kyushu,rcapp_kyushu]{Taikain Suehara}
\author[kek]{Manobu M. Tanaka}
\author[phys_kyushu,rcapp_kyushu]{Junji Tojo}
\author[phys_kyushu]{Keisuke Umebayashi}
\author[phys_kyushu,rcapp_kyushu]{Tamaki Yoshioka}

\affiliation[artsci_kyushu]{organization={Faculty of Arts and Science, Kyushu University},
            addressline={744 Motooka}, 
            city={Fukuoka},
            postcode={819-0395}, 
            country={Japan}}
\affiliation[kek]{organization={Institute of Particle and Nuclear Studies, High Energy Accelerator Research Organization},
                 addressline={1-1 Oho},
                 city={Tsukuba},
                 postcode={305-0801},
                 country={Japan}}
\affiliation[niigata]{organization={Department of Physics, Faculty of Science, Niigata University},
                     addressline={8050 Ikarashi 2-no-cho},
                     city={Niigata},
                     postcode={950-2181},
                     country={Japan}}
\affiliation[toyama]{organization={National Institute of Technology, Toyama College},
                     addressline={1-2 Ebie-neriya},
                     city={Imizu},
                     postcode={933-0293},
                     country={Japan}}
\affiliation[phys_kyushu]{organization={Department of Physics, Faculty of Science, Kyushu University},
                         addressline={744 Motooka},
                         city={Fukuoka},
                         postcode={819-0395},
                         country={Japan}}
\affiliation[rcapp_kyushu]{organization={Research Center for Advanced Particle Physics, Kyushu University},
                          addressline={744 Motooka},
                          city={Fukuoka},
                          postcode={819-0395},
                          country={Japan}}

\begin{abstract}
The SliT ASIC is a readout chip for the silicon strip detector
to be used at the J-PARC muon $g-2$/EDM experiment. The production
version of SliT128D was designed and mass production was finished.
A quality assurance test method for bare SliT128D chips was developed
to provide a sufficient number of chips for the experiment.
The quality assurance 
test of the SliT128D chips was performed and 5735 chips were inspected.
No defect was observed in chips of 84.3\%. Accepting a few channels
with poor time walk performance out of 128 channels per chip, more than 
90\% yield can be achieved, which is sufficient to construct the whole
detector.
\end{abstract}




\end{frontmatter}


\section{Introduction}
\label{sec:introduction}
A silicon strip detector will be used in the J-PARC muon $g-2$/EDM
experiment to detect positrons from muon decay~\cite{e34_exp}.
The detector is composed of 40 vane modules and
each vane module has 16 silicon strip sensors. One sensor has 1024 strips
and the total number of sensor strips in the whole detector 
modules is 655,360.
To readout signals from silicon strip sensors, a new
application specific integrated circuit (ASIC) named SliT was
developed~\cite{SliT128C}. Performance evaluation
was conducted at the pre-production version of SliT128C
and it was confirmed to satisfy all requirements for the
J-PARC muon $g-2$/EDM experiment. The mass-production version
of SliT128D was designed with minor modifications to
SliT1128C  and its mass production was finished.
One SliT128D chip has 128 readout channels so that 5120 chips are
needed to readout all silicon strip sensor signals on the
detector.
To assure normal operation and performance of each 
chip before implementing it on a circuit board, a quality assurance 
system
was developed and a large scale quality assurance test was
conducted to provide sufficient number of chips for the detector.
The test methods and results are described
in this article.

\section{Specifications and requirements of ASIC}
\subsection{Specifications}
The SliT128D chip has a charge-sensitive amplifier (CSA)
for the silicon strip sensor input. Test charge pulses can
be also injected via an AC-coupling capacitor of 100~fF.
The CSA output is fed into a CR-RC shaping amplifier whose
peaking time is about 60~ns.
A differentiator is implemented after the CR-RC
shaper. By using the zero-crossing
timing of the differentiator output corresponding
to the peak time of the CR-RC output,
the time walk of signal can be reduced less than
1~ns which is required from the muon $g-2$/EDM experiment.

Bias voltages and currents for the shaping amplifiers and
common thresholds for the comparators can be
adjusted via pads on an ASIC chip.
The comparator
thresholds for the CR-RC shaper and differentiator
can be adjusted further by 7-bit DACs independently for each channel
by control registers.

The monitor lines are implemented to examine 
analog waveforms of the CSA, CR-RC shaper and differentiator.
Waveforms of one channel selected by a control
register can be monitored.

The signal timing is sampled by an external 200~MHz
frequency clock. When the chip receives ``write start'' 
signal, it starts to store data in SRAMs with
8192 word depth for each channel.

A slow control circuit enables control of
parameters for the analog and digital parts.

\subsection{Requirements for the quality assurance test}
Following functions and performance need to be tested
as quality assurance.

\begin{itemize}
\item Power supply currents and bias voltages
\item Slow control
\item Analog waveforms
\item Digital output of the CR-RC shaper and the differentiator
\item Time walk of the differentiator output
\end{itemize}

\section{Test methods}

\subsection{Probe card}

\begin{figure}[htbp]
  \centering
  \includegraphics[width=0.40\textwidth]{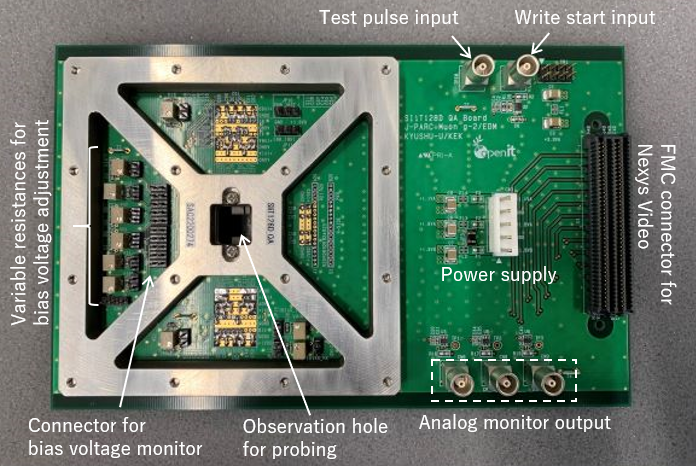}
  \caption{Probe card for the quality assurance test of SliT128D chips.}
  \label{fig:SliT128DProbeCard}
\end{figure}

To test performance of the SliT128D chips, a probe card, which
is a printed circuit board with probe needles, was designed and 
manufactured by Micronics Japan Ltd.
The picture of the probe card is shown in Figure~\ref{fig:SliT128DProbeCard}.

The probe card is a cantilever type and 110 probe needles
made from tungsten-rhenium alloy are implemented.
Probing positions can be viewed from 
the observing hole.
Electric power for a SliT128D chip is supplied
via a connector on the probe card to probe needles. The card also have
analog waveform monitoring output connectors,
and test pulse and write start input connectors.
Variable resistances are implemented on the probe
card for adjustment of bias voltages of a SliT128D chip.
Measured bias voltages can be monitored
from terminals on the probe card by connecting them to the external
voltmeter.

The probe card has a FMC connector to install
a Nexys Video board (manufactured by Digilent Inc.), which is a commercially 
available circuit board for FPGA development.
Digital output is processed in an Artix-7 FPGA on the board.

\subsection{Operation}
\begin{figure}[htbp]
  \centering
  \includegraphics[width=7cm]{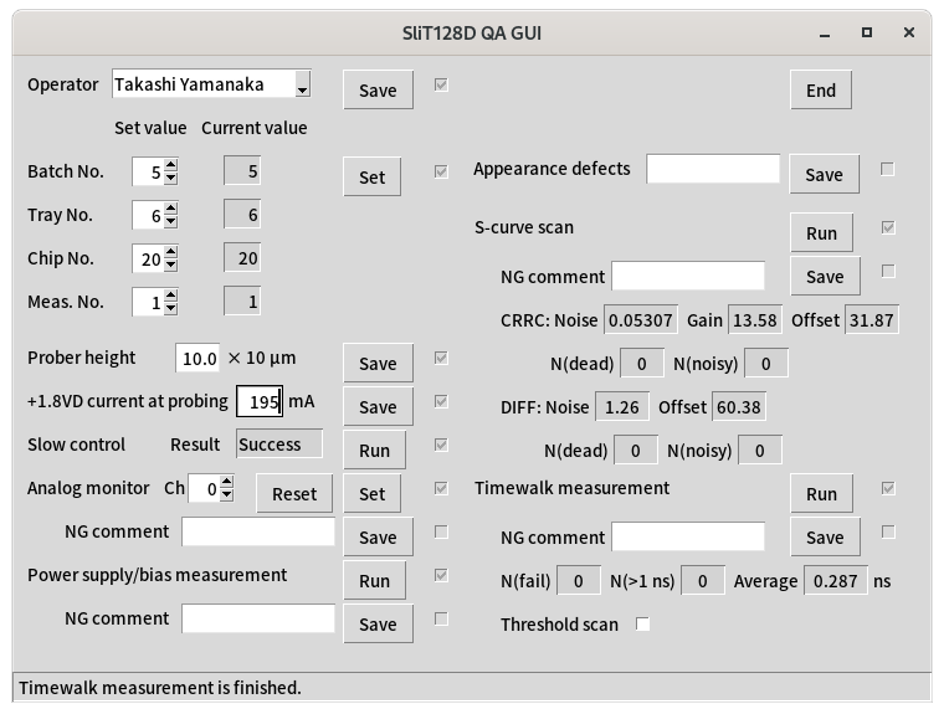}
  \caption{Graphical user interface for the quality 
    assurance test.}
  \label{fig:SliT128D_QA_GUI}
\end{figure}

The quality assurance test was performed with SliT128D chips after dicing.
An operator set one chip on the chuck stage of the probe
system and contacts probe needles of the probe card
to the bonding pads of the chip.
Electric power is supplied to the chip and
measurements are performed one by one using the graphical
user interface shown in Figure~\ref{fig:SliT128D_QA_GUI}.
After finishing the test of a chip, that chip is replaced by
a new chip and the measurements are continued.
A test of one chip typically takes 17 minutes 
including a few minutes of time for chip replacement.

\subsection{Appearance inspection}
Appearance of the chip is inspected by the macroscopy
as well as the microscope.
A chip with any dust or scratch is regarded
as a defective chip. 
It is noted that the most of chips with appearance defects
were removed in the mass production process.

\subsection{Slow control test}

Slow control function of the chip is tested by sending
read and write signal to a control register.
If slow control fails, there is no response from the chip.

\subsection{Analog waveform monitor}

\begin{figure}[htbp]
  \centering
  \includegraphics[width=8cm]{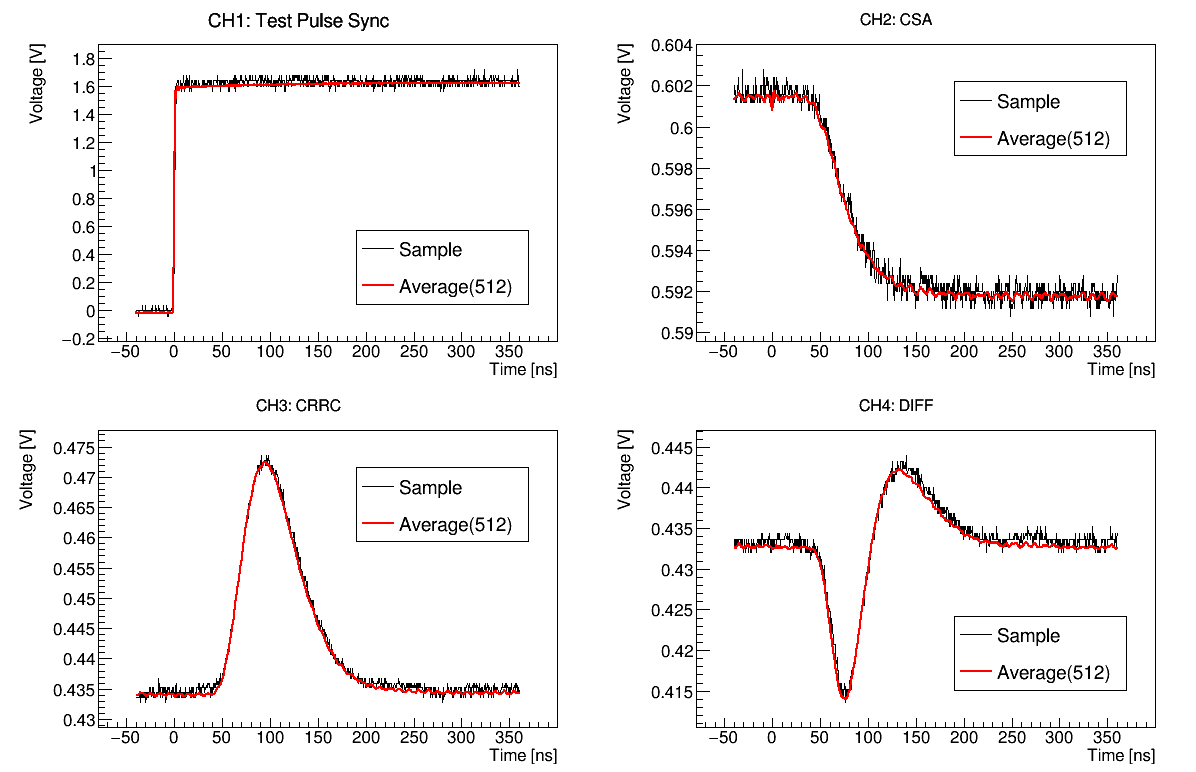}
  \caption{Example of analog waveform monitoring. The CSA (top right),
  the CR-RC (bottom left) and the differentiator (bottom right) waveforms
  are shown with the test pulse timing signal (top left).}
  \label{fig:analog_monitor}
\end{figure}

Analog waveforms of the fixed channel
are obtained by an oscilloscope and are checked
whether they are nominal waveforms or not. Single triggered waveforms and
the average waveforms are taken as shown in
Figure~\ref{fig:analog_monitor}.

\subsection{Power supply currents and bias voltages}

\begin{figure}[htbp]
  \centering
  \includegraphics[width=8cm]{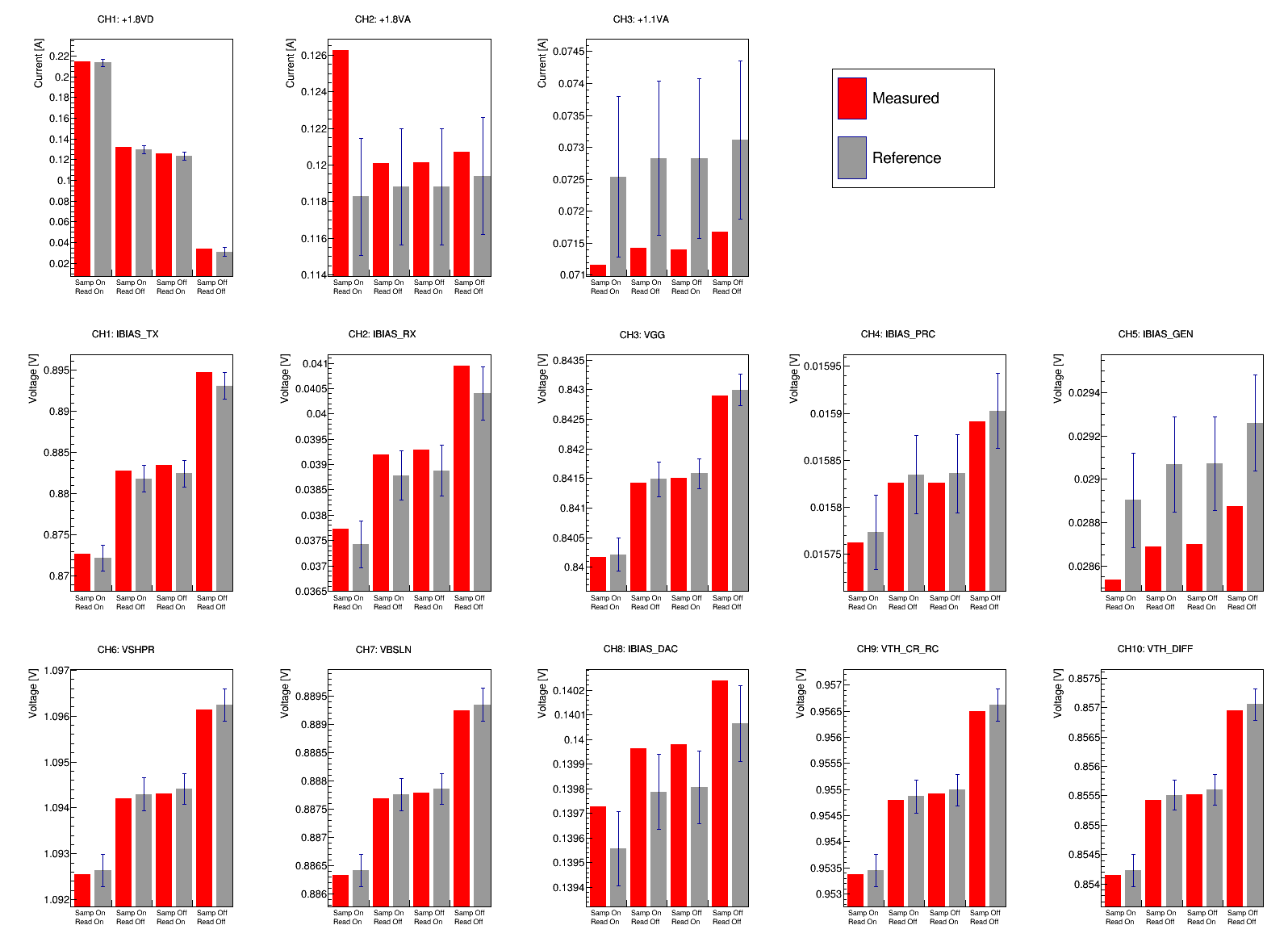}
  \caption{Example of power supply currents and bias voltage
    measurement}
  \label{fig:bias_voltages}
\end{figure}

Power supply currents are readout from a DC power supply.
Bias voltages are monitored via connectors
on the probe card and readout by a multimeter.
These parameters are measured in four different
conditions (whether readout clock and sampling clock
are on or off) and are compared with pre-measured
reference values as shown in Figure~\ref{fig:bias_voltages}.
If measured values significantly deviate from the
reference values, that chip is regarded as defective.

Decrease of the power supply currents and bias voltages
also occur when ASIC pad shavings pile-up on the probe
needles. When significant decrease is observed,
cleaning of probe needles is performed using a wrapping
film sheet and polyurethane cleaning sheet.

\subsection{Threshold scan}
\begin{figure}[htbp]
  \centering
  \includegraphics[width=8.5cm]{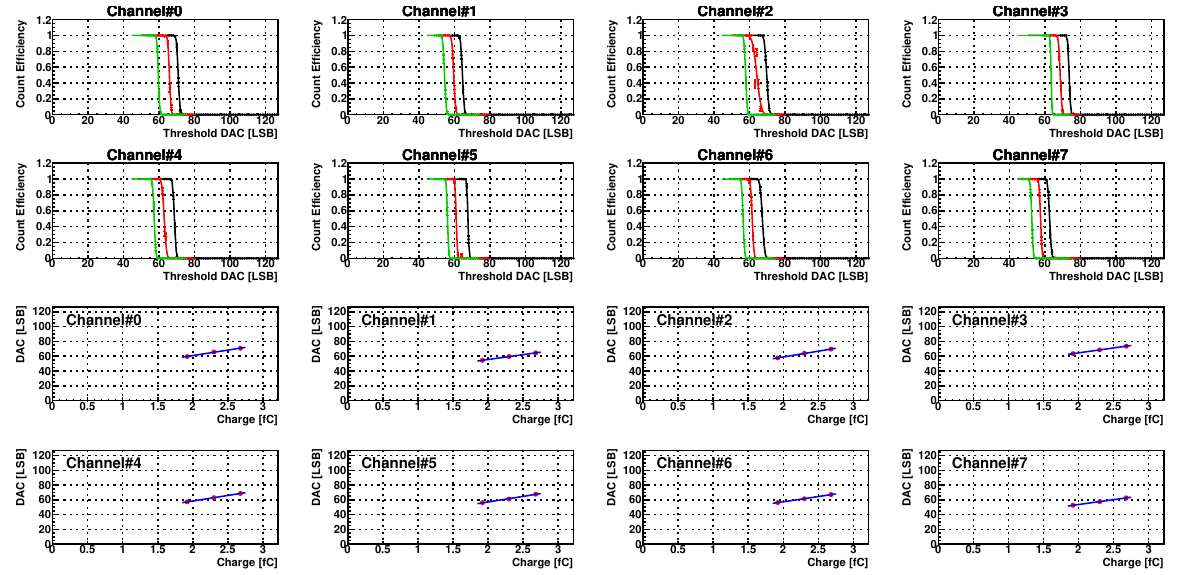}
  \caption{Example of the threshold scan for the CR-RC shaper. Upper eight plots show
  the detection efficiency of test pulses as a function of the comparator threshold.
  Lower eight plots show the center value of the detector efficiency as a function of
  input charges.}
  \label{fig:ThresholdScan}
\end{figure}

Performance of the CR-RC shaper is evaluated by injecting
test charge pulses and counting the detection efficiency
as a function of the comparator threshold.
Three different charges are
injected and noise, gain and baseline voltage
are measured for each channel. Example of the threshold scan for
the CR-RC shaper is shown
in Figure~\ref{fig:ThresholdScan}.
For the differentiator circuit,
the baseline voltage is measured by the threshold scan without injecting
a test charge. 

After these measurements, the threshold of the
CR-RC shaper is set to the nominal value of 0.3 minimum ionizing particle (MIP)
charge and the threshold for the differentiator circuit is set
to the baseline voltage.
Short data is taken with or without a test charge
injection. Noisy channel shows non-existence hit
even without a test charge injection.

\subsection{Time walk measurement}

The time walk performance of the differentiator
circuit is evaluated by injecting test charges
of 0.5~MIP, 1~MIP and 3~MIP to each channel.
The largest signal timing difference between them
is defined as the time walk in the quality assurance test.

The best (smallest) time walk value is obtained when scanning
the threshold for the differentiator circuit for each channel
but it takes too long time for the quality assurance
test. The test with about 600 chips showed
that channels with time walk less than 2~ns
with a fixed threshold near the average of the best thresholds
satisfy the requirement
of 1~ns time walk at the best threshold.
Following this fact, the quality assurance test
in a large volume was performed with the fixed
threshold for time walk measurement.
If the time walk is less than 2~ns, this channel
is regarded to satisfy the requirement.
One chip is sampled out of twenty chips and
the threshold scan is performed for time walk measurement
to confirm this estimate.

\section{Results}
\subsection{Summary of defects}

The test was performed at Kyushu University installing the
probe card on a manual probe system (PM8 manufactured by FormFactor Inc.).
The total number of measured chips was 5735.

Each chip is categorized by the severity
of found defects in the following order.

\begin{enumerate}
\item Appearance defects
\item Abnormal power supply current 
\item Slow control failure
\item Include at least one defective channels
\item Include at least one channel with time walk higher than the requirement
\item Abnormal analog waveforms
\end{enumerate}

If none of the above defects are found, that chip is
regarded as a no defective chip. The test result is
summarized in Table~\ref{tab:result}.

\begin{table}[htbp]
  \caption{The result of the quality assurance test of SliT128D chips. Defects are ordered by the severity.}
  \label{tab:result}
  \begin{center}
    \begin{tabular}{lr}
      \hline
      Defect & Number \\
      \hline
      Appearance defects & 3 \\
      Abnormal power supply current & 23 \\
      Slow control failure & 4 \\
      Defective channels & 291 \\
      High time walk channels & 560 \\
      Abnormal analog waveforms & 21 \\
      No defect & 4833 \\
      \hline
      Total & 5735 \\
      \hline
    \end{tabular}
  \end{center}
\end{table}

\subsection{Appearance defects}
Only three chips were regarded as an appearance defective chip.
They are due to dusts on a chip which are expected to
be attached in manufacturing process, or scratch due
to an operation failure in the quality assurance test.


\subsection{Defective channels}
\begin{figure}[htbp]
  \centering
  \includegraphics[width=8cm]{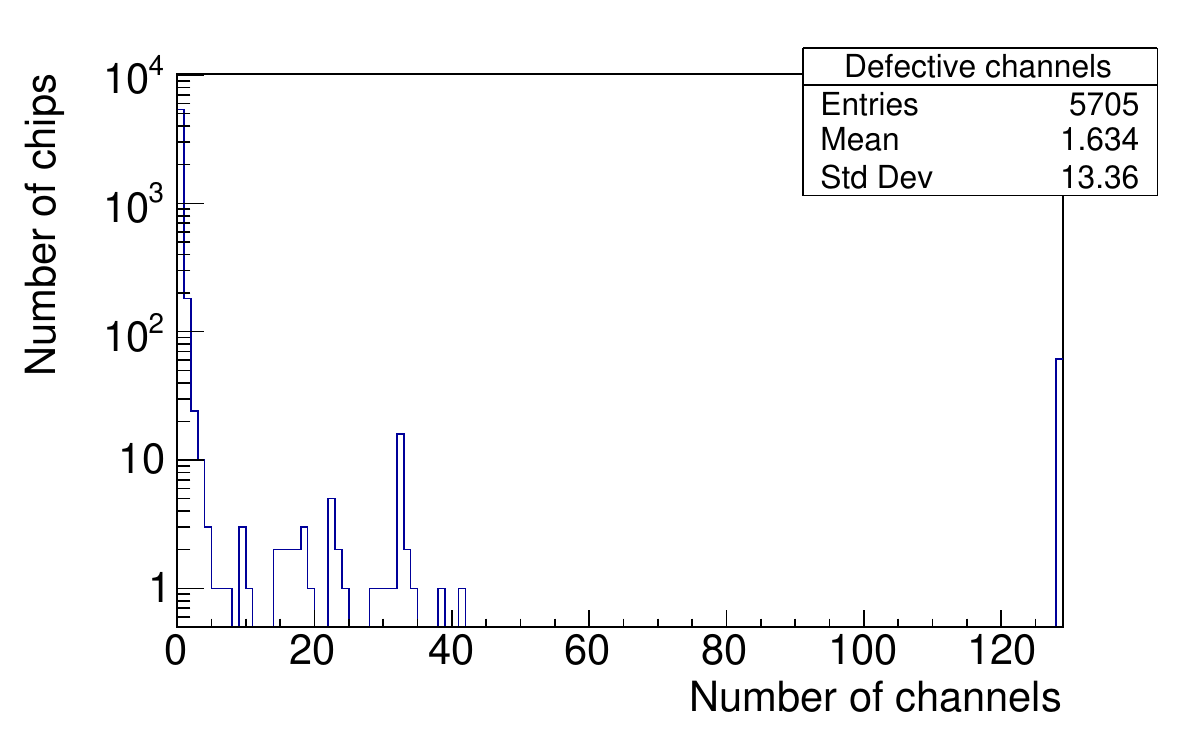}
  \caption{The number of defective channels per chip after removing severer defects.}
  \label{fig:n_defective_channels}
\end{figure}

A channel with no digital output,
having non-existence hits
or showing any other malfunction (e.g. signal inefficiency
even in the lowest comparator threshold, overcounting of signal hits due to ringing waveforms)
is regarded as a defective channel.
The number of defective channels in one chip
after removing chips with severer defects
is shown in Figure~\ref{fig:n_defective_channels}.
The ratio of chips with at least one defective channels
is 5.07\% of the total inspected chips but the most of those chips having only 
a few defective channels. 

\subsection{Time walk measurement}
\begin{figure}[htbp]
  \centering
  \includegraphics[width=8cm]{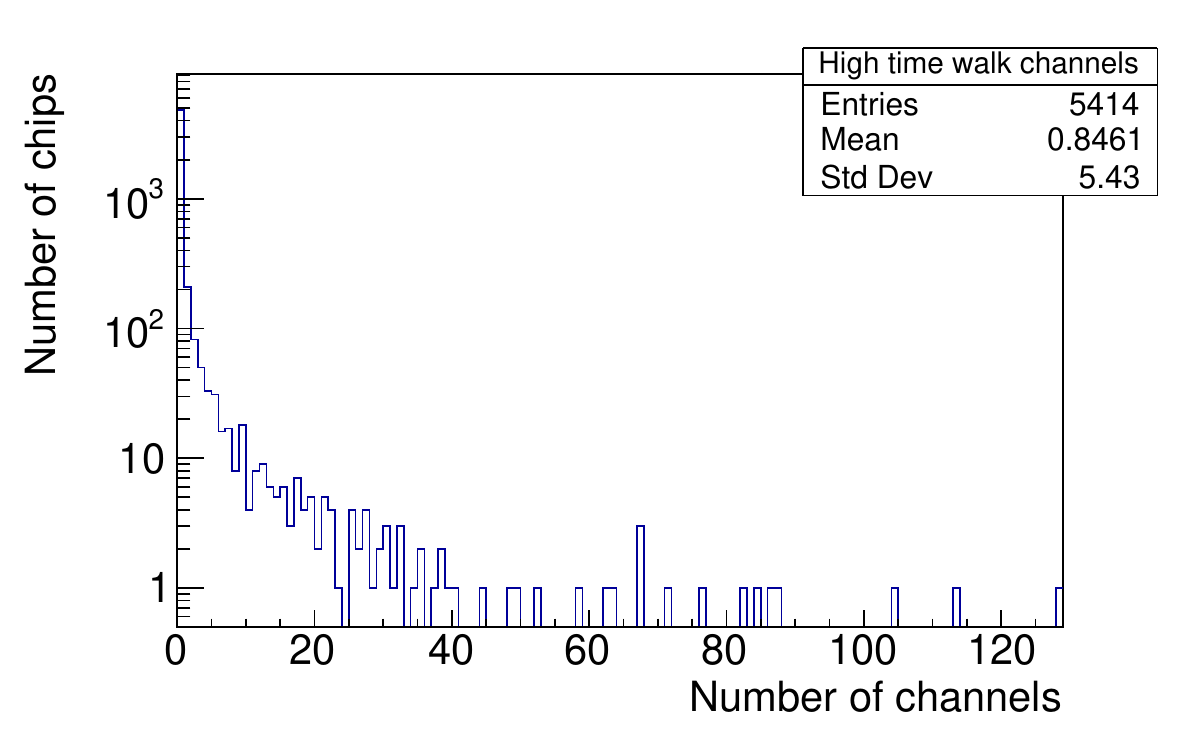}
  \caption{The number of channels per chip whose time walk value
    exceeds the requirement after removing severer defective chips.}
  \label{fig:BadTimewalkChannels}
\end{figure}

The number of channels per chip whose time walk value exceeds
the requirement after removing
chips with severer defects is shown in
Figure~\ref{fig:BadTimewalkChannels}. 
This number distributes up to a lager number
continuously, though the most of chips have only a
few channels with high time walk.
Although channels to be used for hit time measurement
need to satisfy the time walk requirement, it is not necessary that
all channels in one chip satisfy the requirement. Thus, chips with
small number of channels with high time walk
can be used in practice.

\subsection{Abnormal analog waveforms}
Abnormal analog waveforms are observed in 
chips with a certain defect but also in chips without any defects sometimes.
Even though abnormal analog waveform itself does not affect practical usage 
if digital output is fine, 
those chips are also regarded as defective chips for safety.

\subsection{Yield}
\begin{figure}[htbp]
  \centering
  \includegraphics[width=9cm]{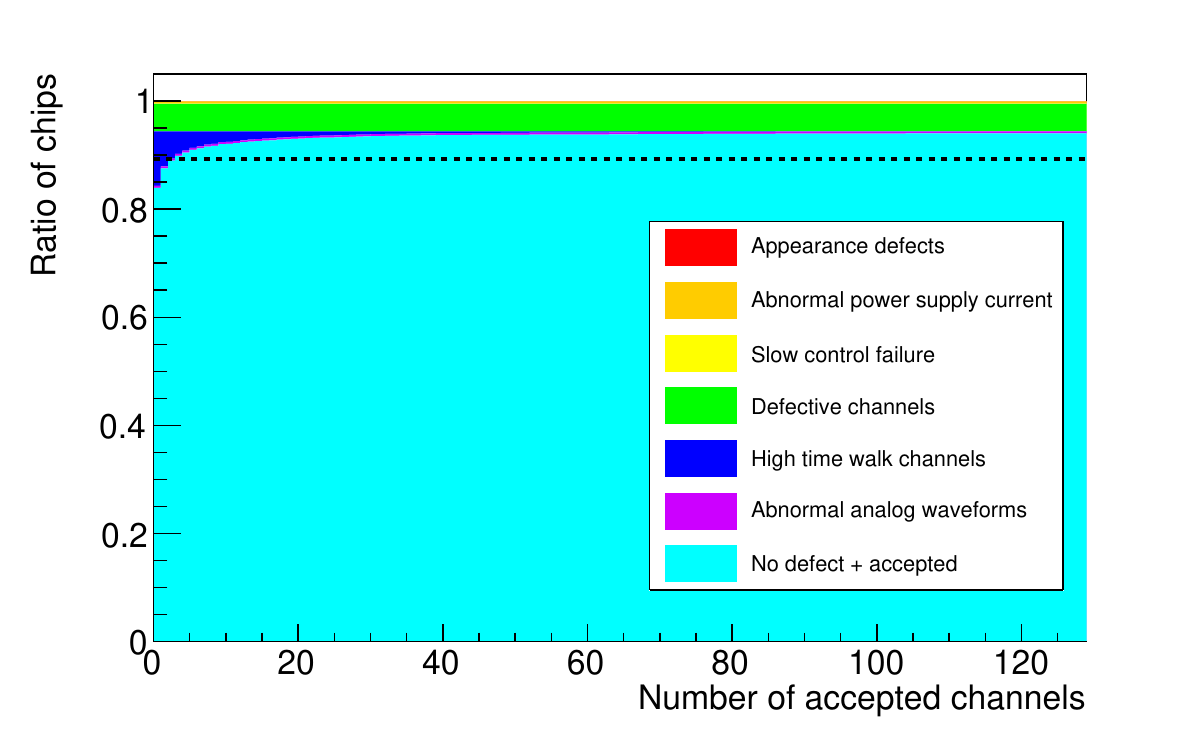}
  \caption{The ratio of defective chips as a function of the number of accepted
    channels with high time walk. A dotted line indicates the target yield of
  89.3\%.}
  \label{fig:final_yield}
\end{figure}

The ratio of no defective chips is 84.3\%.
To provide a sufficient number of chips for the experiment (i.e. 5120 chips),
more than 89.3\% yield is needed.
This yield can be obtained by accepting a few channels with
high time walk in one chip.
The yield as a function of the number of accepted channels with
high time walk is shown in Figure~\ref{fig:final_yield}.
To secure 5120 chips, at least three channels with high time walk need to
be accepted.

\section{Conclusion}

The quality assurance test of SliT128D chips was performed
to confirm their normal operation and performance before assembling into
the detector module for the muon $g-2$/EDM experiment
at J-PARC. The test of 5735 chips were performed and
84.3\% chips were found to have no defects. The most of
defects are the existence of at least one defective channels
or channels with high time walk than the requirement. Accepting
at least three channels with high time walk in one chip,
a sufficient number of chips for the whole detector
can be provided.

\appendix


\section*{Acknowledgments}
This research was supported by JSPS KAKENHI Grants No.
JP20H05625 and 22H01232. The authors would like to 
acknowledge the Open Source Consortium of Instrumentation (Open-It)
of KEK for their support on the electronics design.

\bibliographystyle{elsarticle-num} 
\bibliography{references}





\end{document}